\documentclass[twocolumn,showpacs,preprintnumbers,amsmath,amssymb,prb]{revtex4}
\usepackage{graphicx}
\usepackage{dcolumn}
\usepackage{bm}

\begin{document}

\preprint{APS/123-QED}

\title{Critical behavior of repulsive dimers on square lattices at $2/3$ monolayer coverage}

\author{F. Rom\'a}
\affiliation{Centro At{\'{o}}mico Bariloche, R8402AGP San Carlos
de Bariloche, R\'{\i}o Negro, Argentina}
\affiliation{Departamento
de F\'{\i}sica, Universidad Nacional de San Luis, Chacabuco 917,
D5700BWS San Luis, Argentina}
\author{J. L. Riccardo}
\affiliation{Departamento de F\'{\i}sica, Universidad Nacional de
San Luis, Chacabuco 917, D5700BWS San Luis, Argentina}
\author{A.J. Ramirez-Pastor}
\affiliation{Departamento de F\'{\i}sica, Universidad Nacional de
San Luis, Chacabuco 917, D5700BWS San Luis, Argentina}

\date{\today}

\begin{abstract}
Monte Carlo simulations and finite-size scaling theory have been
used to study the critical behavior of repulsive dimers on square
lattices at $2/3$ monolayer coverage. A ``zig-zag" (ZZ) ordered
phase, characterized by domains of parallel ZZ strips oriented at
$\pm 45^o$ from the lattice symmetry axes, was found. This ordered
phase is separated from the disordered state by a order-disorder
phase transition occurring at a finite critical temperature. Based
on the strong axial anisotropy of the ZZ phase, an orientational
order parameter has been introduced. All the critical quantities
have been obtained. The set of critical exponents suggests that
the system belongs to a new universality class.
\end{abstract}

\pacs{68.35.Rh, 64.60.Cn, 68.43.De, 05.10.Ln}

\maketitle

\section{Introduction}

The study of critical phenomena and phase transitions is a major
and long standing topic in statistical physics.
~\cite{Stanley,Fisher,Kawasaki,Baxter,Yeomans,Goldenfeld,Domb}
Particularly, the two-dimensional lattice-gas model~\cite{Hill}
with repulsive interactions between the adparticles has received
considerable theoretical and experimental interest because it
provides the theoretical framework to study structural
order-disorder transitions occurring in many adsorbed monolayer
films.~\cite{Dash,Taub,Somorjai,Schick1,Binder0,Binder1,Landau1,Binder2,Landau2,Landau3,Schick2,Schick3,Schick4,Patrykiejew}
Most studies have been devoted to adsorption of particles with
single occupancy. The problem becomes considerably difficult when
particles occupy two adjacent lattice sites (dimers).
Consequently, there have been a few studies devoted to
order-disorder transitions associated to dimer adsorption with
repulsive lateral interactions. Among them, the structural
ordering of interacting dimers has been analyzed by A. J. Phares
et al.~\cite{Phares} The authors calculated the entropy of dimer
on semi-infinite $M \times N$ square lattice ($N \rightarrow
\infty$) by means of transfer matrix techniques. They concluded
that there are a finite number of ordered structures. As it arose
from simulation analysis,~\cite{SURFSCI3} only two of the
predicted structures survive at thermodynamic limit. In fact, in
Ref.~\onlinecite{SURFSCI3}, the analysis of the phase diagram for
repulsive nearest-neighbor interactions on a square lattice
confirmed the presence of two well-defined structures: a $c(4
\times 2)$ ordered phase at $\theta=1/2$ and a ``zig-zag" (ZZ)
order at $\theta = 2/3$, being $\theta$ the surface coverage.

The thermodynamic implication of such a structural ordering was
demonstrated through the analysis of adsorption
isotherms,~\cite{LANG5} the collective diffusion
coefficient~\cite{SURFSCI2} and the configurational
entropy~\cite{LANG6} of dimers with nearest-neighbor repulsion.
Later, Monte Carlo (MC) simulations and finite-size scaling (FSS)
techniques have been used to study the critical behavior of
repulsive linear $k$-mers in the low-coverage ordered structure
(at $\theta=1/2$).~\cite{PRB4,PRB5} A ($2k \times 2$) ordered
phase, characterized by alternating lines, each one being a
sequence of adsorbed $k$-mers separated by $k$ adjacent empty
sites, was found. The critical temperature and critical exponents
were calculated. The results revealed that the system does not
belong to the universality class of the two-dimensional Ising
model. The study was extended to triangular lattices.~\cite{PRB6}
In this case, the exponents obtained for $k>1$ and
$\theta=k/(2k+1)$ are very close to those characterizing the
critical behavior of $k$-mers ($k>1$) on square lattices at
$\theta=1/2$.

Recently, by using MC simulations and finite-scaling techniques,
R\.zysko and Bor\'owko have studied a wide variety of systems in
presence of
multisite-occupancy.~\cite{BORO1,BORO2,BORO3,BORO5,BORO4} Among
them, attracting dimers in the presence of energetic
heterogeneity;~\cite{BORO1} heteronuclear dimers consisting of
different segments A and B adsorbed on square
lattices;~\cite{BORO2,BORO3,BORO5} and trimers with different
structures adsorbed on square lattices.~\cite{BORO4} In these
leading papers, a rich variety of phase transitions was reported
along with a detailed discussion about critical exponents and
universality class.

Summarizing, although there have been various studies for
monolayers at half coverage, to the author's knowledge, there are
no conclusive studies on the characteristics of the transition
phase of repulsive dimers on a square lattice at $2/3$ coverage.
In the present contribution we attempt to remedy this situation.
For this purpose, extensive MC simulations in the canonical
ensemble complemented by analysis using FSS techniques have been
applied. The FSS study has been divided in two parts. Namely, $1)$
a conventional FSS in terms of the normalized scaling variable
$\epsilon \equiv T/T_c -
1$,~\cite{Fisher,Binder0,Privman,Privman1} where $T_c$ is the
critical temperature; and $2)$ an extended FSS,
~\cite{GARTENHAUS,CAMPBELL} where $\sigma \equiv 1-T_c/T$, instead
of $\epsilon$, is used. Our results led the determination of the
critical temperature separating the transition between a
disordered state and the ZZ ordered phase occurring at $2/3$
coverage and the critical exponents characterizing the phase
transition.

The outline of the paper is as follows: In Sec. II we describe the
dimer lattice-gas model. The order parameter and the simulation
scheme are introduced in Secs. III and IV, respectively. Finally,
the results and general conclusions are presented in Sec. V.

\section{The model}

In this section, the lattice-gas model for dimer adsorption is
described. The surface is represented as a simple square lattice
in two-dimensions consisting of $M=L \times L$ adsorptive sites,
where $L$ is the size of the system along each axis. The
homonuclear dimer is modelled as $2$ monomers at a fixed
separation, which equals the lattice constant $a$. In the
adsorption process, it is assumed that each monomer occupies a
single adsorption site and the admolecules adsorb or desorb as one
unit, neglecting any possible dissociation. The high-frequency
stretching motion along the molecular bond has not been considered
here.

\begin{figure}[t]
\includegraphics[width=4cm,clip=true,angle=90]{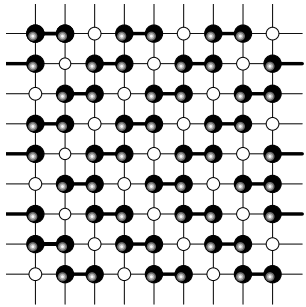}
\caption{Snapshot of the ordered phase for dimers at $\theta
=2/3$.} \label{figure1}
\end{figure}

In order to describe the system of $N$ dimers adsorbed on $M$
sites at a given temperature $T$, let us introduce the occupation
variable $c_i$ which can take the following values: $c_i=0$ if the
corresponding site is empty and $c_i = 1$ if the site is occupied.
Under this consideration, the Hamiltonian of the system is given
by,
\begin{equation}
H = w \sum_{\langle i,j \rangle} c_i c_j - N w+ \epsilon_o
\sum_{i} c_i \label{h}
\end{equation}
where $w$ is the nearest-neighbor (NN) interaction constant which
is assumed to be repulsive (positive),$\langle i,j \rangle$
represents pairs of NN sites and $\epsilon_o$ is the energy of
adsorption of one given surface site. The term $Nw$ is subtracted
in Eq.~(\ref{h})  since the summation over all the pairs of NN
sites overestimates the total energy by including $N$ bonds
belonging to the $N$ adsorbed dimers.

\section{Order parameter}

Given the inherent anisotropy of the adparticles, it is convenient
to define a related order parameter. In this section, we will
briefly refer to a recently reported order parameter
$\delta$,~\cite{PRB5} which measures the orientation of the
admolecules in the ordered structure.

Fig.~\ref{figure1} shows one of the possible configurations of the
ordered ZZ structure appearing for dimers at $2/3$ monolayer.
Though the degeneracy of this phase is high, the entropy per
lattice site tends to zero in the thermodynamic
limit~\cite{LANG6}. The figure suggests a simple way to build an
order parameter. In fact, any realization of the ZZ structure
implies the orientation of the particles along one of the lattice
axis.~\cite{foot1} Then, all the available configurations can be
grouped in two sets, according to this orientation. Taking
advantage of this property, we define the order parameter as:
\begin{equation}
\delta =   \left | \frac{N_v-N_h}{N}  \right | \label{fi2}
\end{equation}
where $N_v$ ($N_h$) represents the number of dimers aligned along
the vertical (horizontal) axis and $N= N_v + N_h$.

When the system is disordered $(T>T_c)$, the two orientations
(vertical or horizontal) are equivalent and $\delta$ is zero. As
the temperature is decreased below $T_c$, the dimers align along
one direction and $\delta$ is different from zero. Thus, $\delta$
appears as a proper order parameter to elucidate the phase
transition.

\section{Monte Carlo method}

The lattices were generated fulfilling the following conditions:

\begin{itemize}
\item[1)] The sites were arranged in a square lattice of side $L$
($M = L \times L$), with conventional periodic boundary
conditions. \item[2)] Because the surface was assumed to be
homogeneous, the interaction energy between the adsorbed dimer and
the atoms of the substrate $\epsilon_o$  was neglected for sake of
simplicity. \item[3)] In order to maintain the lattice at $2/3$
coverage, $\theta=2N/M=2/3$, the number of dimers on the lattice
was fixed as $N = M/3$. \item[4)] Appropriate values of $L$ ($=60,
72, 84, 96, 108$) were used in such a way that the ZZ adlayer
structure is not altered by boundary conditions.
\end{itemize}

In order to study the critical behavior of the system, we have
used an exchange MC method.~\cite{Hukushima,Earl} As in Ref.
~\onlinecite{Hukushima}, we build a compound system that consists
of $m$ noninteracting replicas of the system concerned. The $i$-th
replica is associated with a heat bath at temperature $T_i$ [or
$\beta_i=1/(k_B T_i)$, $k_B$ being the Boltzmann constant]. To
determine the set of temperatures, $\{T_i \}$, we set the highest
temperature, $T_1$, in the high-temperature phase where relaxation
(correlation) time is expected to be very short and there exists
only one minimum in the free energy space. On the other hand, the
lowest temperature, $T_m$, is set in the low-temperature phase
whose properties we are interested in. Finally, the difference
between two consecutive temperatures, $T_i$ and $T_{i+1}$ with
$T_i > T_{i+1}$, is set as $\Delta T = \left(T_{1} - T_m
\right)/(m-1)$ (equally spaced temperatures).

Under these conditions, the algorithm to carry out the simulation
process is built on the basis of two major subroutines: {\it
replica-update} and {\it exchange}.

\noindent {\it Replica-update}:  Interchange vacancy-particle and
diffusional relaxation. The procedure is as follows: (a) One out
of the $m$ replicas is randomly selected (for example the $i$-th
replica). (b) A dimer and a pair of nearest-neighbor empty sites,
both belonging to the replica chosen in (a), are randomly selected
and their coordinates are established. Then, an attempt is made to
interchange its occupancy state with probability given by the
Metropolis rule, ~\cite{Metropolis}:
\begin{equation}
P = \min \left\{1,\exp\left( - \beta_i \Delta H \right) \right\}
\end{equation}
where $\Delta H$ is the difference between the Hamiltonians of the
final and initial states. (c) A dimer is randomly selected. Then,
a displacement is attempted (following the Metropolis scheme), by
either jumps along the dimer axis or reptation through a $90^o$
rotation of the dimer axis, where one of the dimer centers remains
in its position (interested readers are referred to Fig.~ 1 in
Ref. ~\onlinecite{SURFSCI2} for a more complete description of the
reptation mechanism). This procedure (diffusional relaxation) must
be allowed in order to reach equilibrium in a reasonable time.

\noindent {\it Exchange}: Exchange of two configurations $X_i$ and
$X_{i'}$, corresponding to the $i$-th and $i'$-th replicas,
respectively, is tried and accepted with probability
$W\left(X_i,\beta_i| X_{i'},\beta_{i'}\right)$. In general, the
probability of exchanging configurations of the $i$-th and $i'$-th
replicas is given by, ~\cite{Hukushima}
\begin{equation}
W\left(X_i,\beta_i| X_{i'},\beta_{i'}\right)=\left\{
\begin{array}{cc}
1 & {\rm for}\ \ {\Delta \leq 0} \\
\exp(-\Delta)  & {\rm for}\ \ {\Delta>0}
\end{array}
\right.
\end{equation}
where $\Delta=\left( \beta_i - \beta_{i'} \right)\left[ H(X_{i'})
- H(X_{i}) \right]$. As in Ref. ~\onlinecite{Hukushima}, we
restrict the replica-exchange to the case $i'= i+1$.

The complete simulation procedure is the following:

\begin{itemize}

\item[1)] Initialization.

\item[2)] Replica-update.

\item[3)] Exchange.

\item[4)] Repeat from step 2) $m \times M $ times. This is the
elementary step in the simulation process or Monte Carlo step
(MCS).

\end{itemize}

The initialization of the compound system of $m$ replicas, step
1), is as follows.  By starting with a random initial condition,
the configuration of the replica $1$ is obtained after $n_1$
MCS$'$ at $T_1$ (MCS$'$ consists of $M$ realizations of the
replica-update subroutine). Second, for $i=\{ 2,....,m \}$, the
configuration of the $i$-th replica  is obtained after $n_1$
MCS$'$ at $T_i$, taking as initial condition the configuration of
the replica to $T_{i-1}$.  This method results more efficient than
a random initialization of each replica.

The procedure 1)-4) is repeated for all lattice sizes. For each
lattice, the equilibrium state can be well reproduced after
discarding the first $n_2$ MCS. Then, averages are taken over
$n_{MCS}$ successive MCS. As it was mentioned above, a set of
equally spaced temperatures is chosen in order to accurately
calculate the physical observables in the close vicinity of $T_c$.

The thermal average $\langle ... \rangle$ of a physical quantity
$A$ is obtained through simple averages:
\begin{equation}
{\langle A\rangle}=\frac{1}{n_{MCS}} \sum_{t=1}^{n_{MCS}} A
\left[X_i(t)\right].
\end{equation}
In the last equation, $X_i$ stands for the state of the $i$-th
replica (at temperature $T$). Thus, the specific heat $C$ (in
$k_B$ units) is sampled from energy fluctuations:
\begin{equation}
\frac{C}{k_B}= \frac{1}{(L k_B T)^2} [\langle H^2 \rangle -
\langle H \rangle ^2].
\end{equation}
The quantities related with the order parameter, such as the
susceptibility $\chi$, and the reduced fourth-order cumulant $U$
introduced by Binder, ~\cite{Binder2} are calculated as:
\begin{equation}
\chi = \frac{L^2}{k_BT} [ \langle \delta^2 \rangle - \langle
\delta \rangle^2]
\end{equation}
and
\begin{equation}
U = 1 -\frac{\langle \delta^4\rangle} {3\langle
\delta^2\rangle^2}. \label{cum}
\end{equation}

Finally, in order to discuss the nature of the phase transition,
the fourth-order energy cumulant, $U_E$, was obtained as:
\begin{equation}
U_E=1 -\frac{\langle H^4\rangle} {3\langle H^2 \rangle^2}
\label{cume}
\end{equation}

\begin{table}
\caption{Parameters used in the MC runs.  For all lattice sizes we
have chosen $T_1=0.25$ and $T_m=0.15$ (the temperatures are in
units of $w/k_B$).}
\begin{ruledtabular}
\begin{tabular}{ccccc}
$L$ & $m$ & $n_1$ & $n_2$ & $n_{MCS}$  \\
\hline
$60$  & $101$ & $10^5$       & $10^5$        & $3$ x $10^5$     \\
$72$  & $121$ & $10^5$       & $10^5$        & $3$ x $10^5$     \\
$84$  & $141$ & $2$ x $10^5$ & $2$ x $10^5$  & $5$ x $10^5$     \\
$96$  & $161$ & $4$ x $10^5$ & $4$ x $10^5$  & $8$ x $10^5$     \\
$108$ & $141$ & $5$ x $10^5$ & $5$ x $10^5$  & $10^6$     \\
\end{tabular}
\end{ruledtabular}
\end{table}

\section{Results and conclusions}

The critical behavior of the present model has been investigated
by means of the computational scheme described in the previous
section and FSS techniques.  The values of the parameters used in
each MC run are shown in Table I. In addition, all the simulation
calculations were obtained by averaging over $50$ MC runs.

Because the replica temperatures were chosen equally spaced, the
acceptance probability of the replica-exchange decreases in the
critical temperature region, reaching a minimum whose value is
always greater than $50 \%$. The equilibration has been tested by
studying how the results vary when the simulation times $n_2$ and
$n_{MCS}$ are successively increased by factors of $2$.  We
require that the last three results for all observables agree
within error bars. This simple method is shown to be useful to
test equilibration [see, for instance, Ref.~\onlinecite{KATZ}].
All calculations were carried out using the parallel cluster BACO
of Universidad Nacional de San Luis, Argentina. This facility
consists of 60 PCs each with a 3.0 GHz Pentium-4 processor.

\begin{figure}
\includegraphics[width=6cm,clip=true]{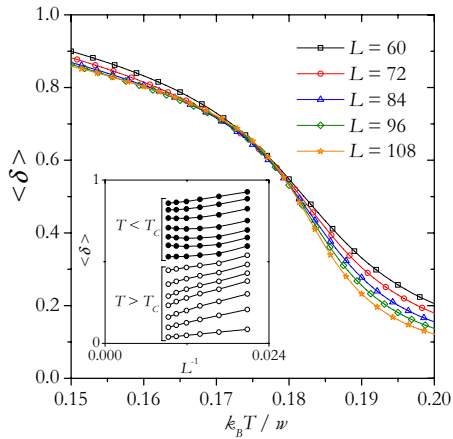}
\caption{(Color online) Size dependence of the order parameter as
a function of temperature. Inset: dependence of $\langle \delta
\rangle$ on $L^{-1}$ for different regimes of $T$ as indicated.
The error in each measurement is smaller that the size of the
symbols.} \label{figure2}
\end{figure}

\begin{figure}
\includegraphics[width=6cm,clip=true]{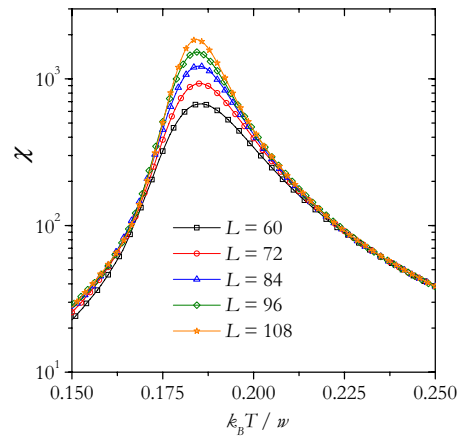}
\caption{(Color online) Size dependence of the susceptibility as a
function of temperature. The error in each measurement is smaller
that the size of the symbols.} \label{figure3}
\end{figure}

\begin{figure}
\includegraphics[width=6cm,clip=true]{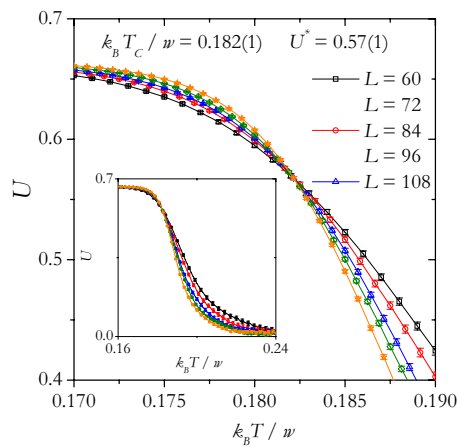}
\caption{(Color online) $U$ versus $k_BT/w$, for different sizes.
From their intersections one obtained $k_BT_c/w$. In the inset,
the data are plotted over a wider range of temperatures.}
\label{figure4}
\end{figure}

\begin{figure}
\includegraphics[width=6cm,clip=true]{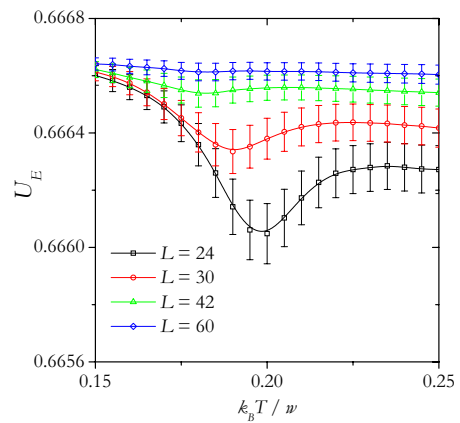}
\caption{(Color online) Temperature variation of $U_E$ for various
lattice sizes.} \label{figure5}
\end{figure}

The conventional FSS implies the following behavior of $C$,
$\langle \delta \rangle$, $\chi$ and $U$ at criticality,
\begin{equation}
C=L^{\alpha/\nu} \tilde C(L^{1/\nu} \epsilon) \label{cal}
\end{equation}
\begin{equation}
\langle \delta \rangle = L^{-\beta/\nu} \tilde \delta(L^{1/\nu}
\epsilon) \label{phi}
\end{equation}
\begin{equation}
\chi= L^{\gamma/\nu}\tilde \chi(L^{1/\nu} \epsilon) \label{chi}
\end{equation}
\begin{equation}
U=\tilde U(L^{1/\nu} \epsilon) \label{cum}
\end{equation}
\noindent for $L \rightarrow \infty$, $\epsilon \rightarrow 0$
such that $L^{1/\nu} \epsilon $= finite, where ($ \epsilon \equiv
T/T_c - 1$). Here $\alpha$, $\beta$, $\gamma$ and $\nu$ are the
standard critical exponents of the specific heat ( $C \sim
|\epsilon|^{-\alpha}$ for $\epsilon \rightarrow 0$, $L \rightarrow
\infty $), order parameter ($\langle \delta \rangle \sim
-\epsilon^{\beta} $ for $\epsilon\rightarrow 0^-$, $L\rightarrow
 \infty$),
 susceptibility($\chi \sim |\epsilon|^\gamma
$ for $\epsilon \rightarrow 0$, $L\rightarrow \infty$) and
correlation length $\xi$ ($\xi \sim |\epsilon|^{-\nu}$ for
$\epsilon \rightarrow 0, L \rightarrow \infty$), respectively.
$\tilde C, \tilde \delta, \tilde \chi $ and $\tilde U$ are scaling
functions for the respective quantities. In the case of extended
FSS, ~\cite{GARTENHAUS,CAMPBELL} $\epsilon$ and $L$ in
eqs.~(\ref{cal}-\ref{cum}) are replaced by $\sigma \equiv 1-T_c/T$
and $L\sqrt{T}$, respectively.

\begin{figure}[t]
\includegraphics[width=5.5cm,clip=true]{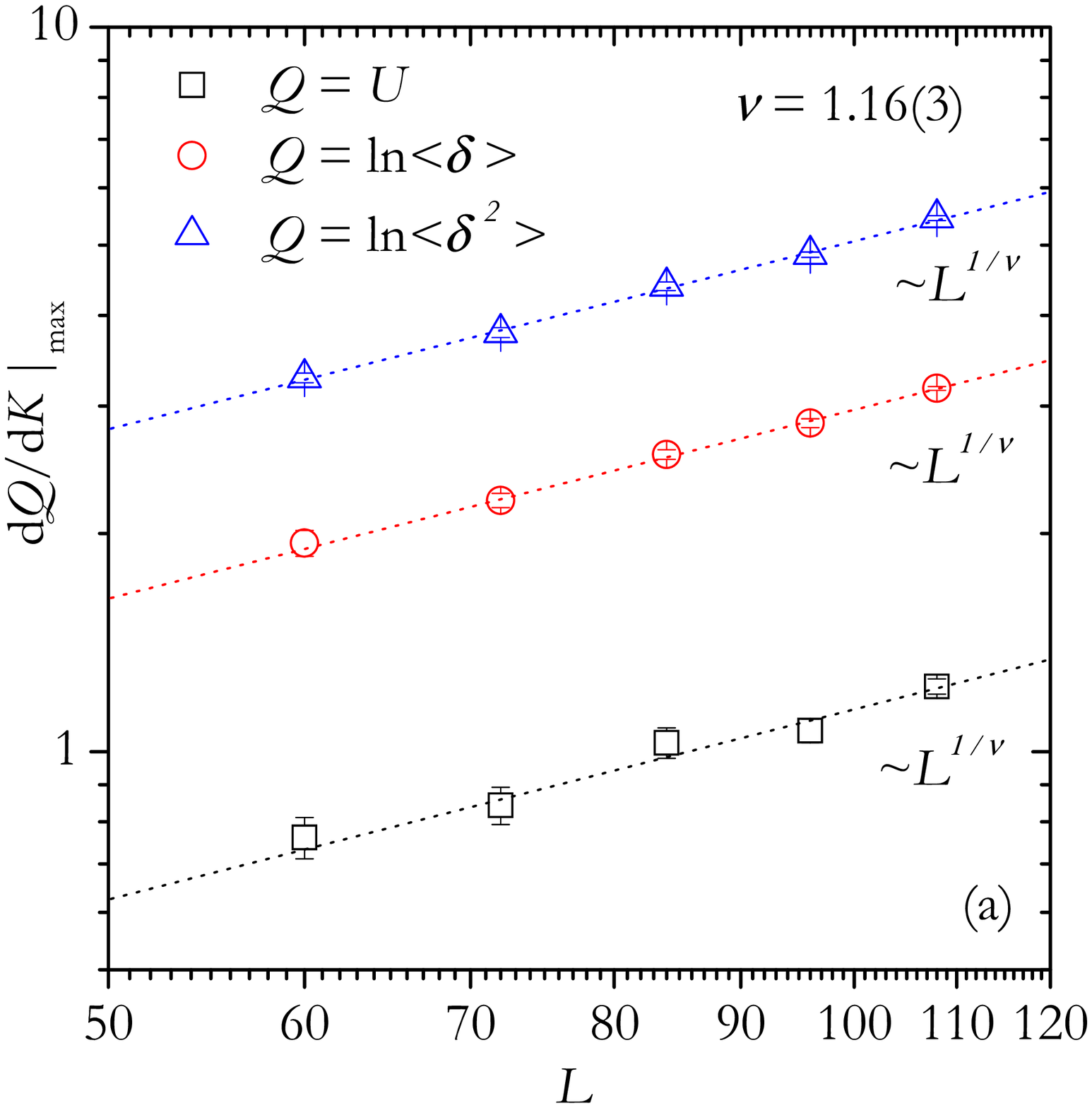}
\includegraphics[width=5.5cm,clip=true]{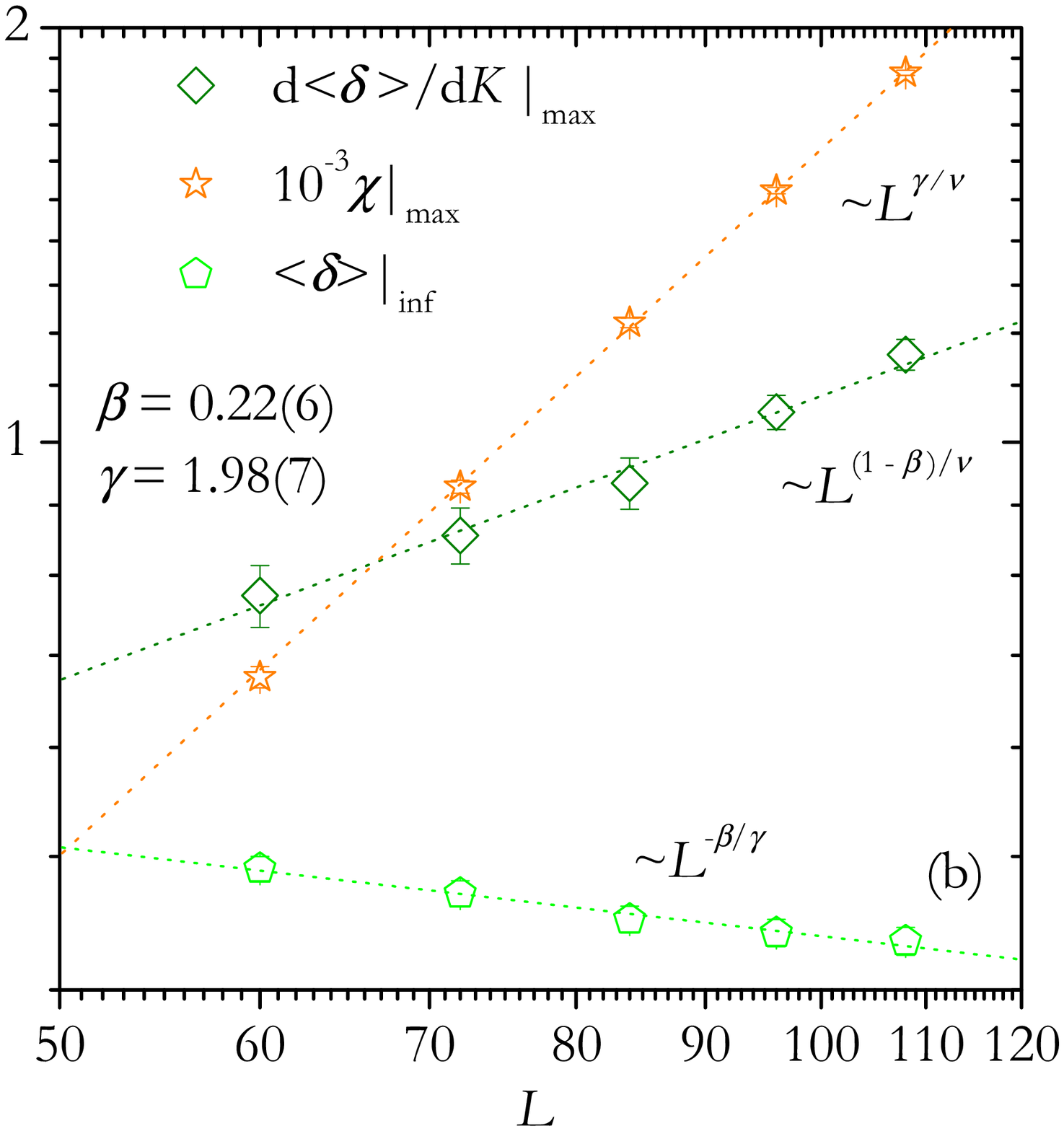}
\includegraphics[width=5.5cm,clip=true]{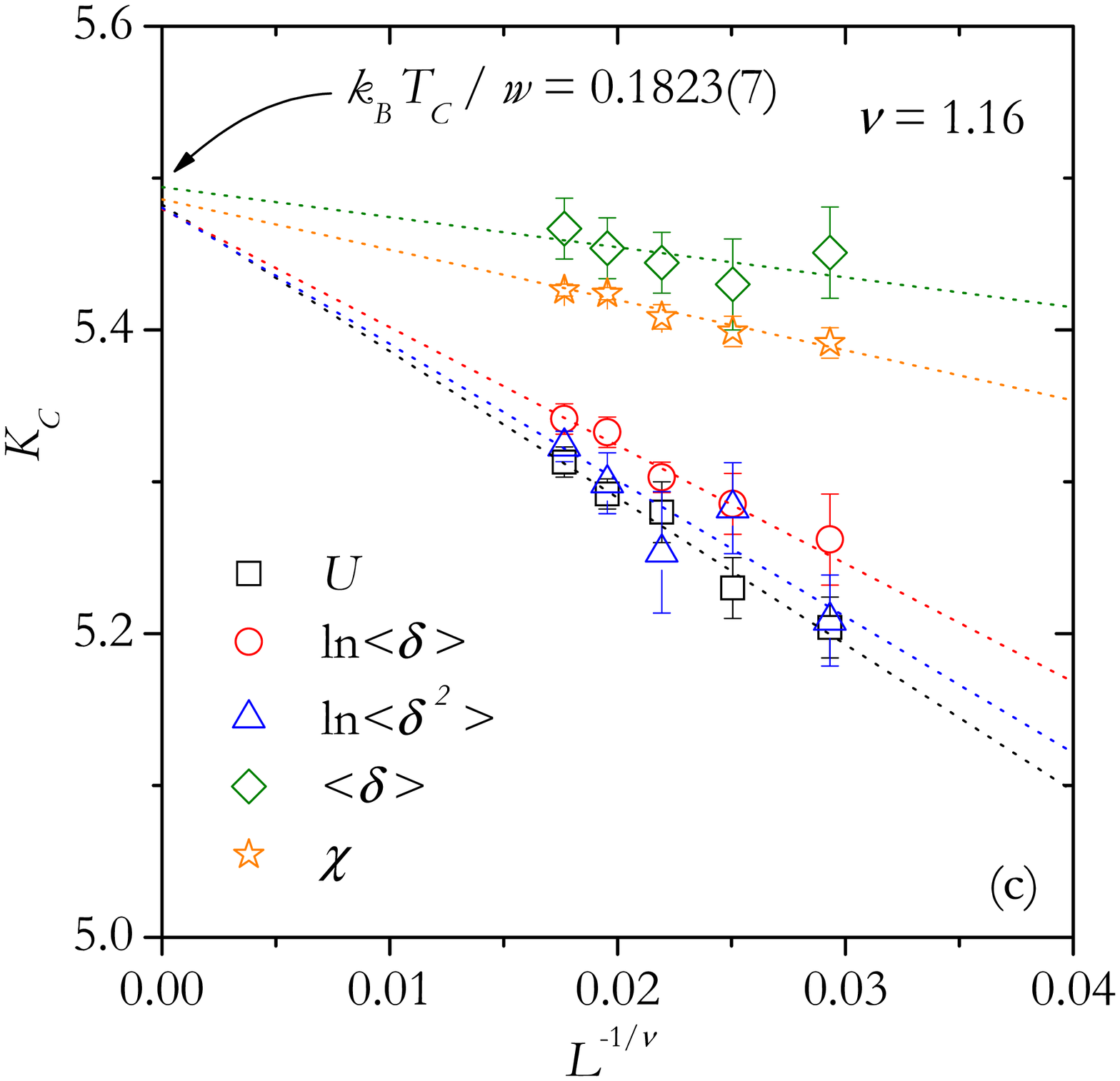}
\caption{(Color online) (a) Log-log plot of the size dependence of
the maximum values of derivatives of various thermodynamic
quantities used to determine $\nu$.  (b) Log-log plot of the size
dependence of the maximum value of the susceptibility, the point
of inflection of the order parameter and the maximum value of the
derivative of the order parameter used to determine $\gamma$ and
$\beta$, respectively. (c) $K_c(L)$ vs. $L^{-1/\nu}$ from several
quantities as indicated. From extrapolation one obtains the
estimation of the critical temperature. In all cases, dotted lines
correspond to linear fits of the data and $L=60, 72,84,96,108$. }
\label{figure6}
\end{figure}

\begin{figure}
\includegraphics[width=5.8cm,clip=true]{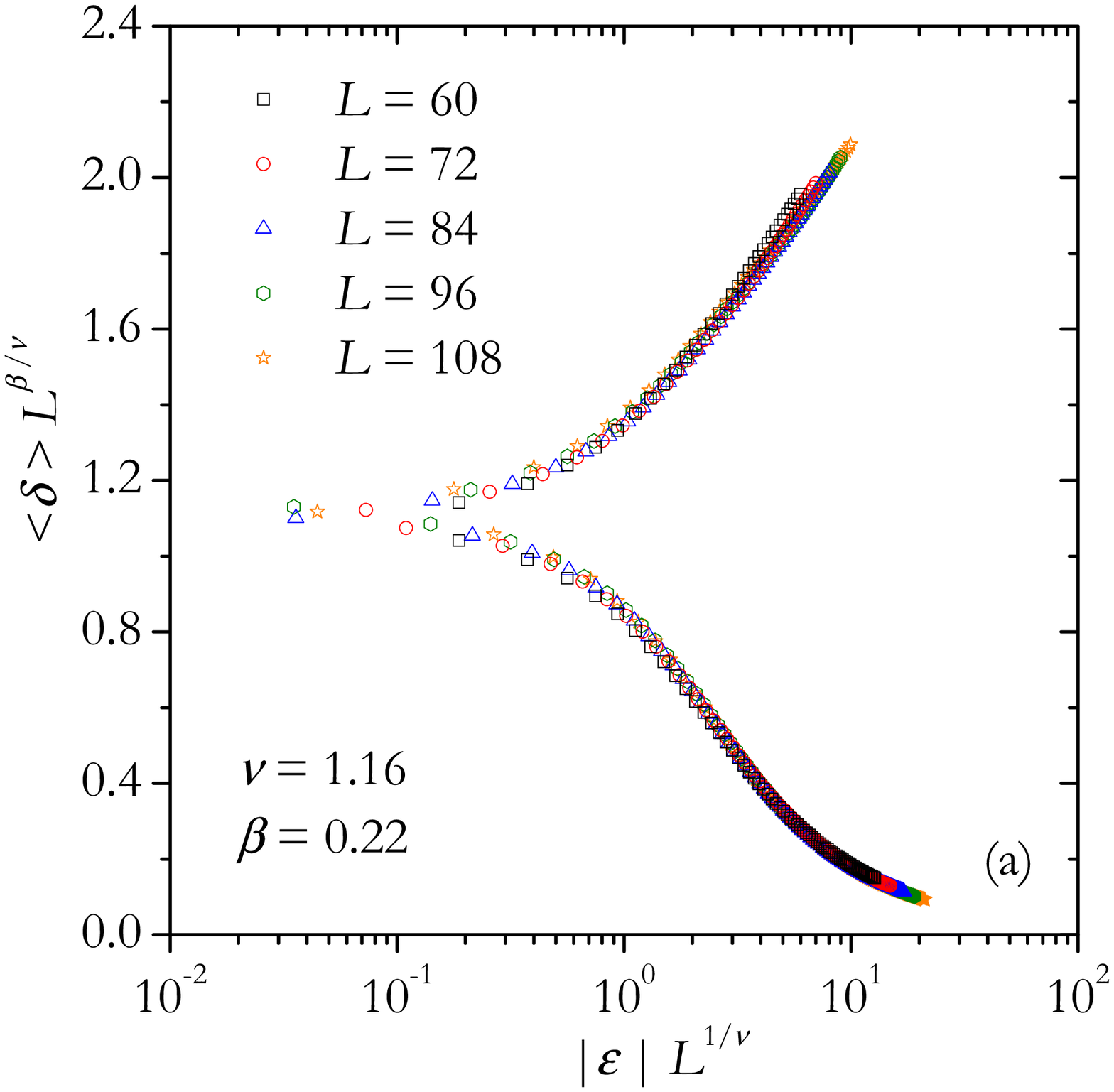}
\includegraphics[width=5.8cm,clip=true]{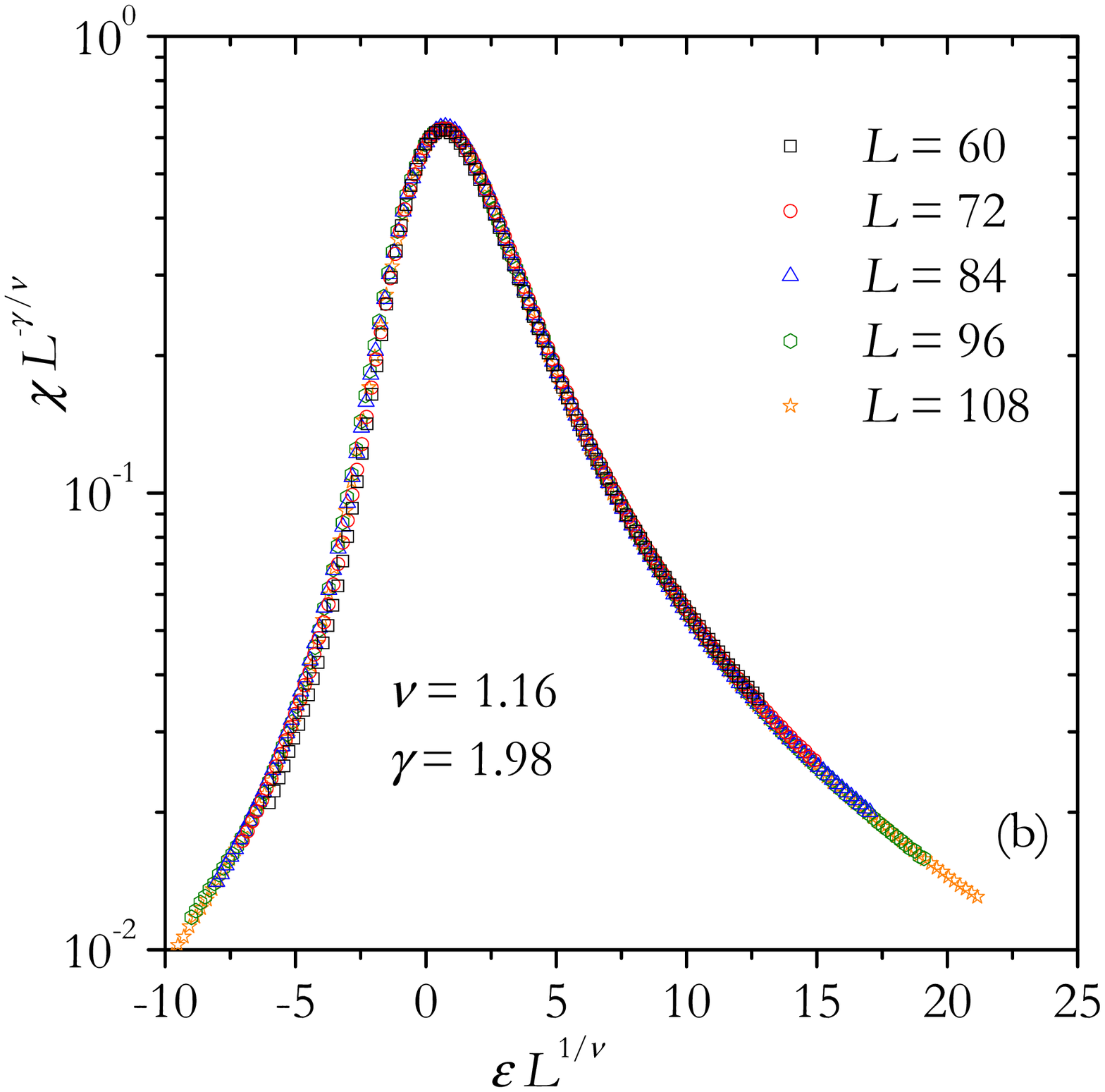}
\includegraphics[width=5.8cm,clip=true]{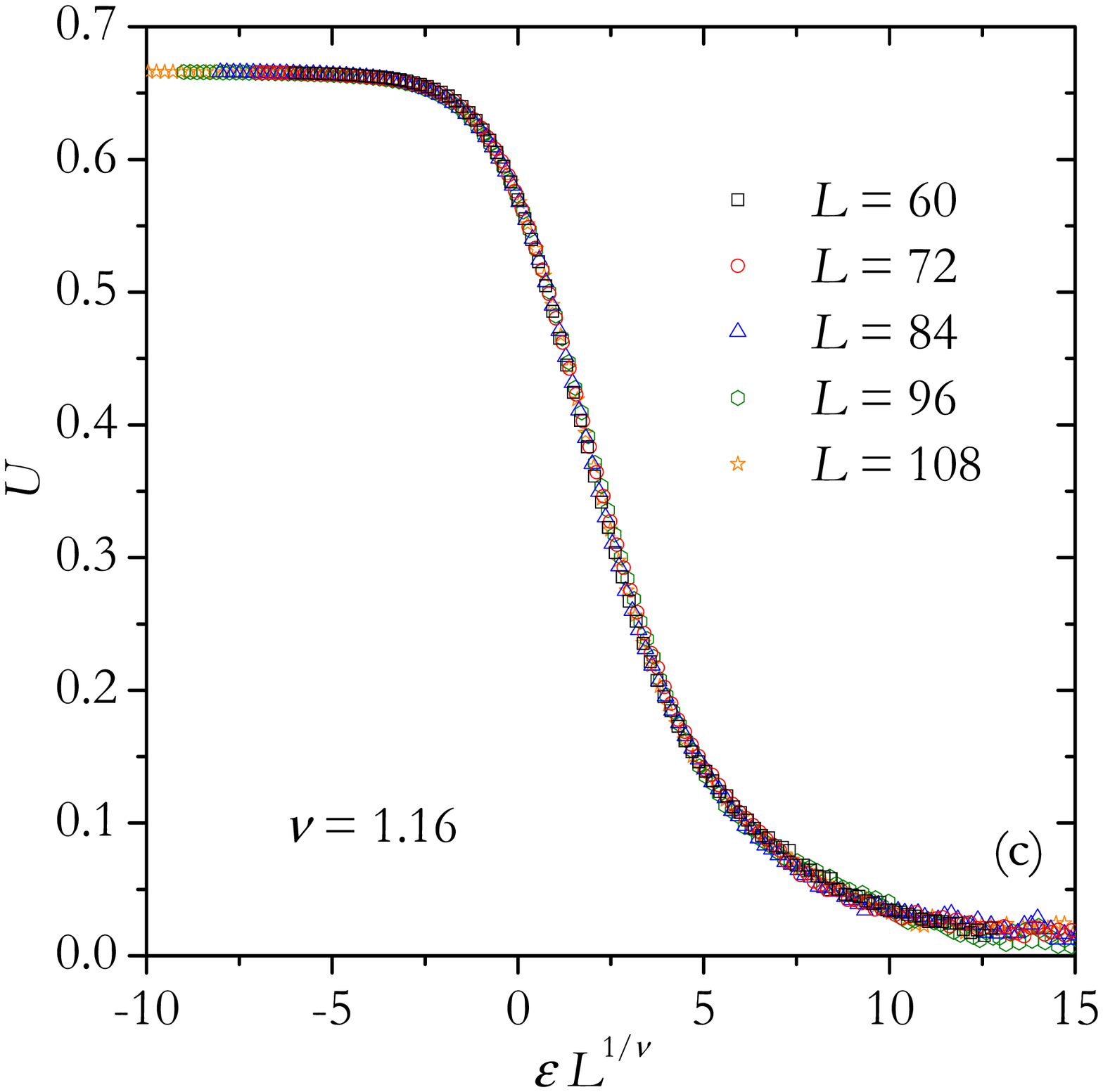}
\caption{(Color online) Conventional data collapsing for: (a) the
curves in Fig.~\ref{figure2}; (b) the curves in
Fig.~\ref{figure3}; and (c) the curves in Fig.~\ref{figure4}.}
\label{figure7}
\end{figure}

\begin{figure}
\includegraphics[width=5.8cm,clip=true]{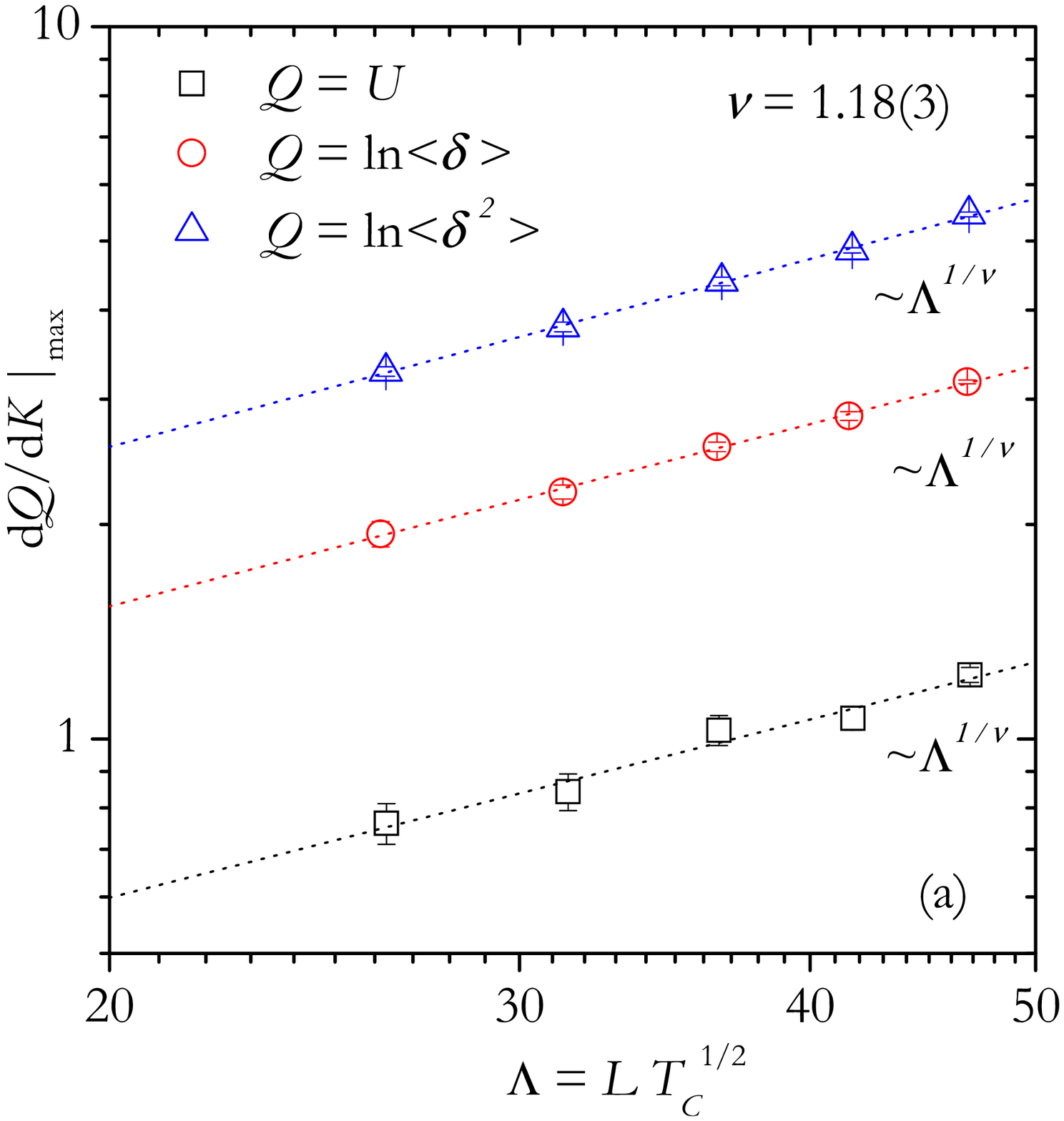}
\includegraphics[width=5.8cm,clip=true]{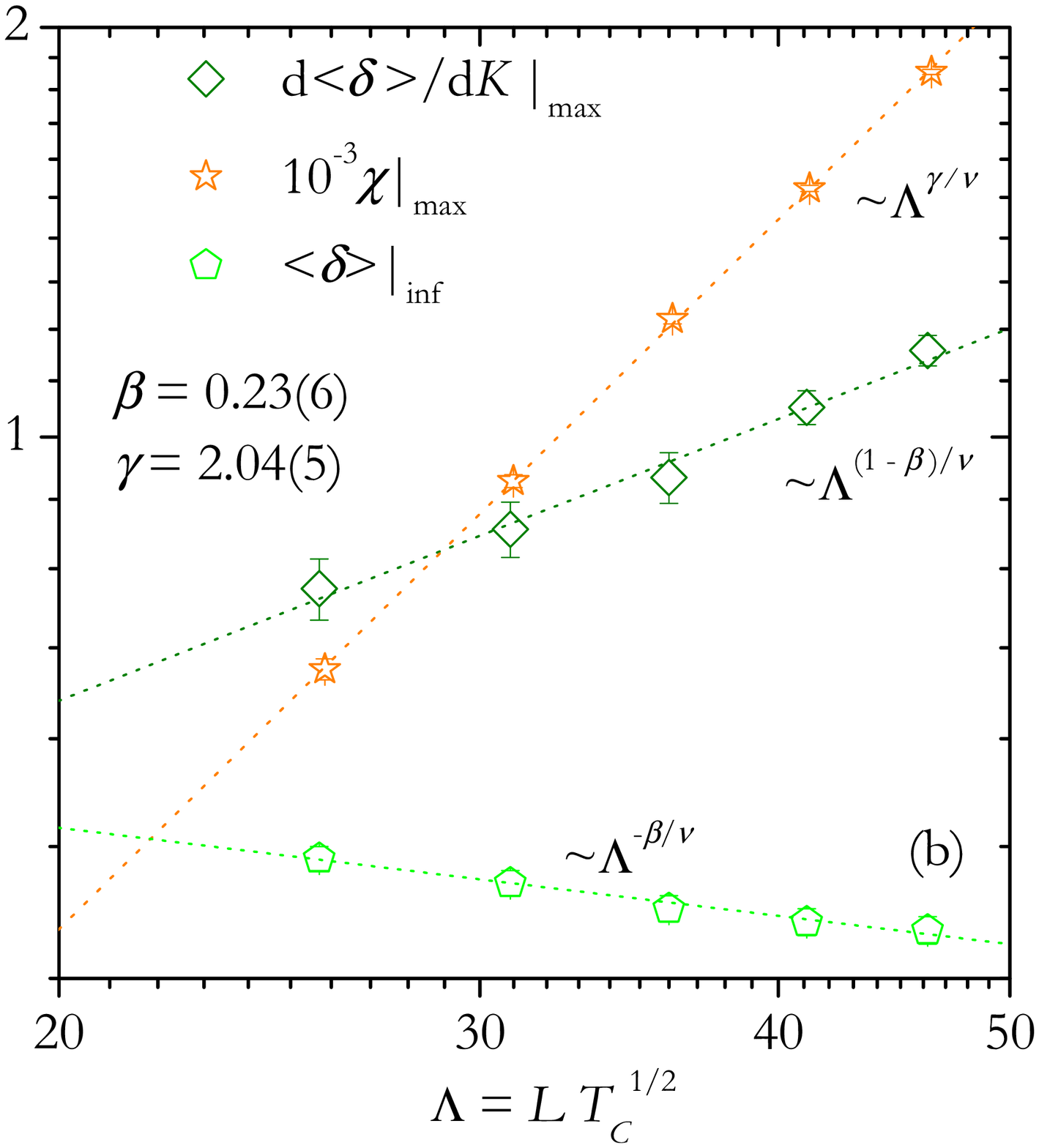}
\includegraphics[width=5.8cm,clip=true]{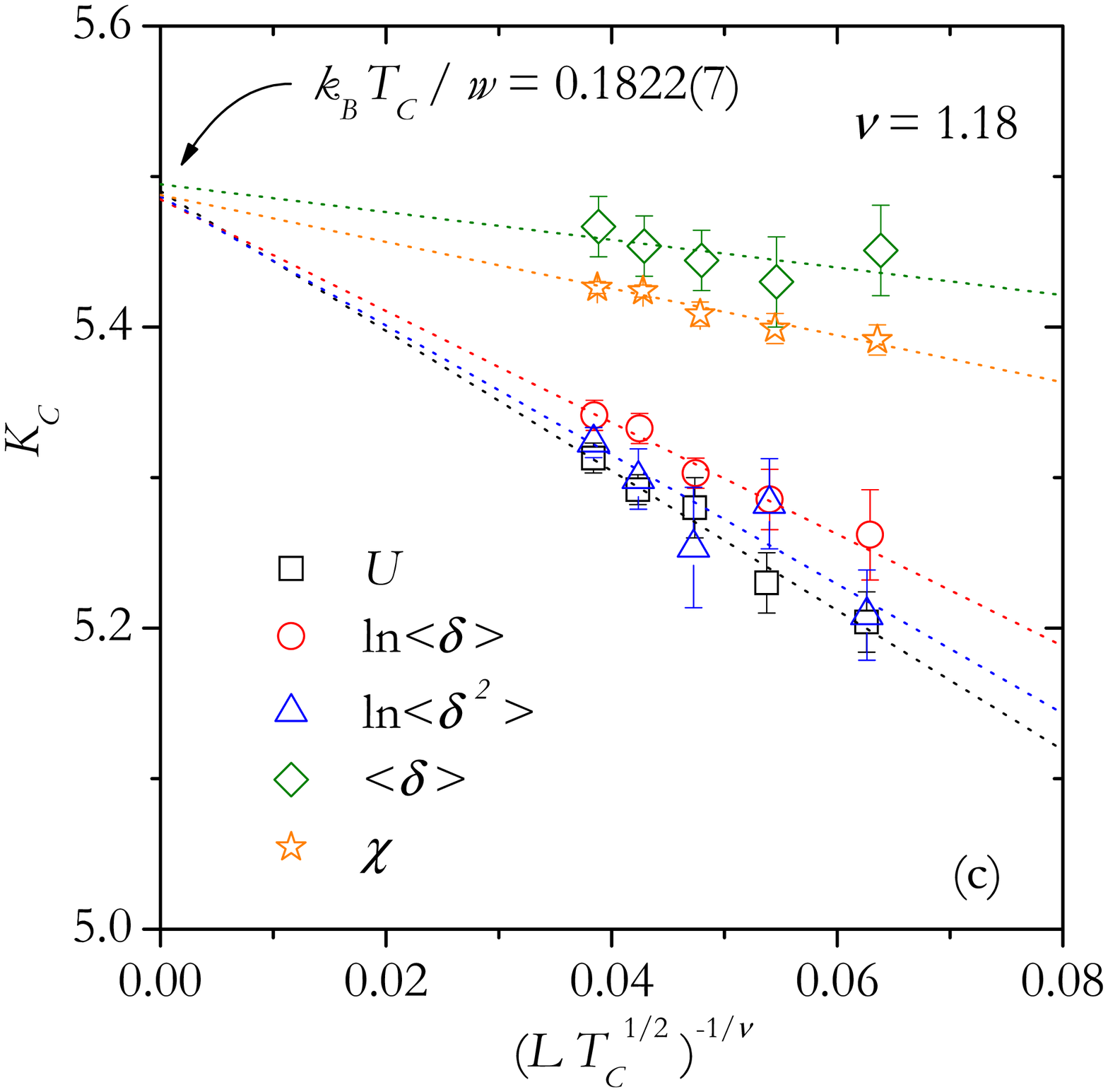}
\caption{(Color online) Same as Fig.~\ref{figure6} for the
extended scaling scheme.} \label{figure8}
\end{figure}

\begin{figure}
\includegraphics[width=5.8cm,clip=true]{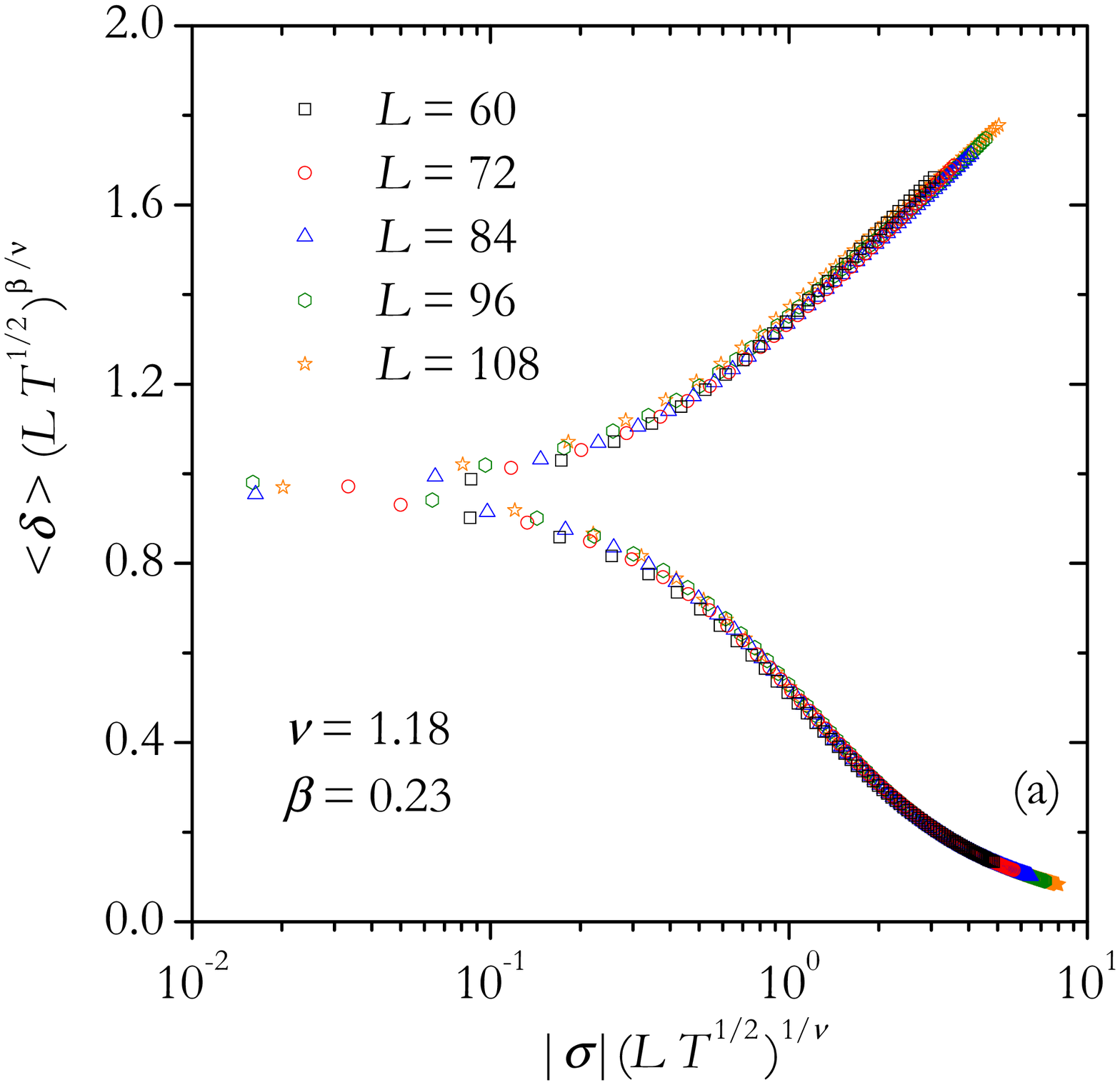}
\includegraphics[width=5.8cm,clip=true]{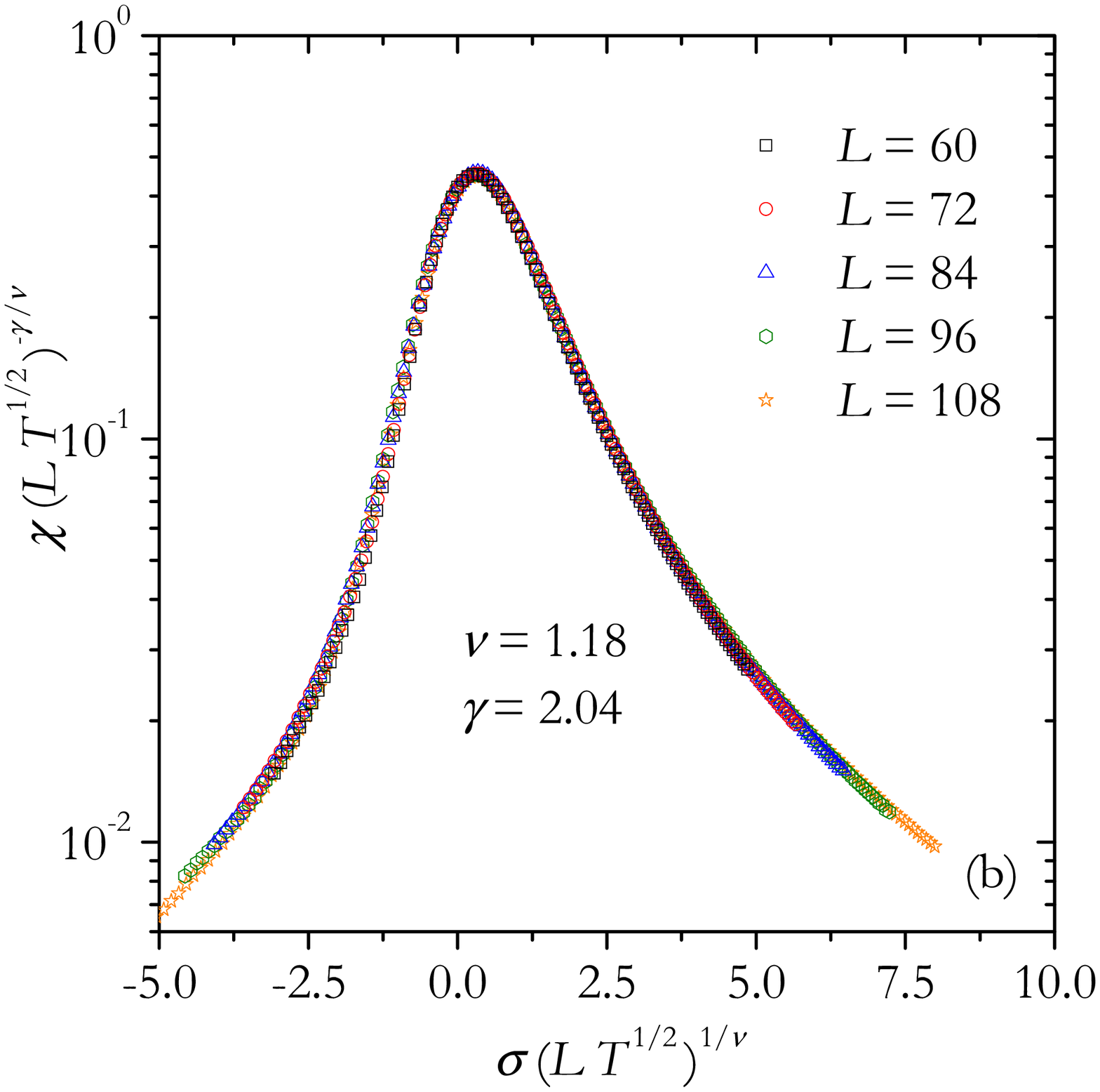}
\includegraphics[width=5.8cm,clip=true]{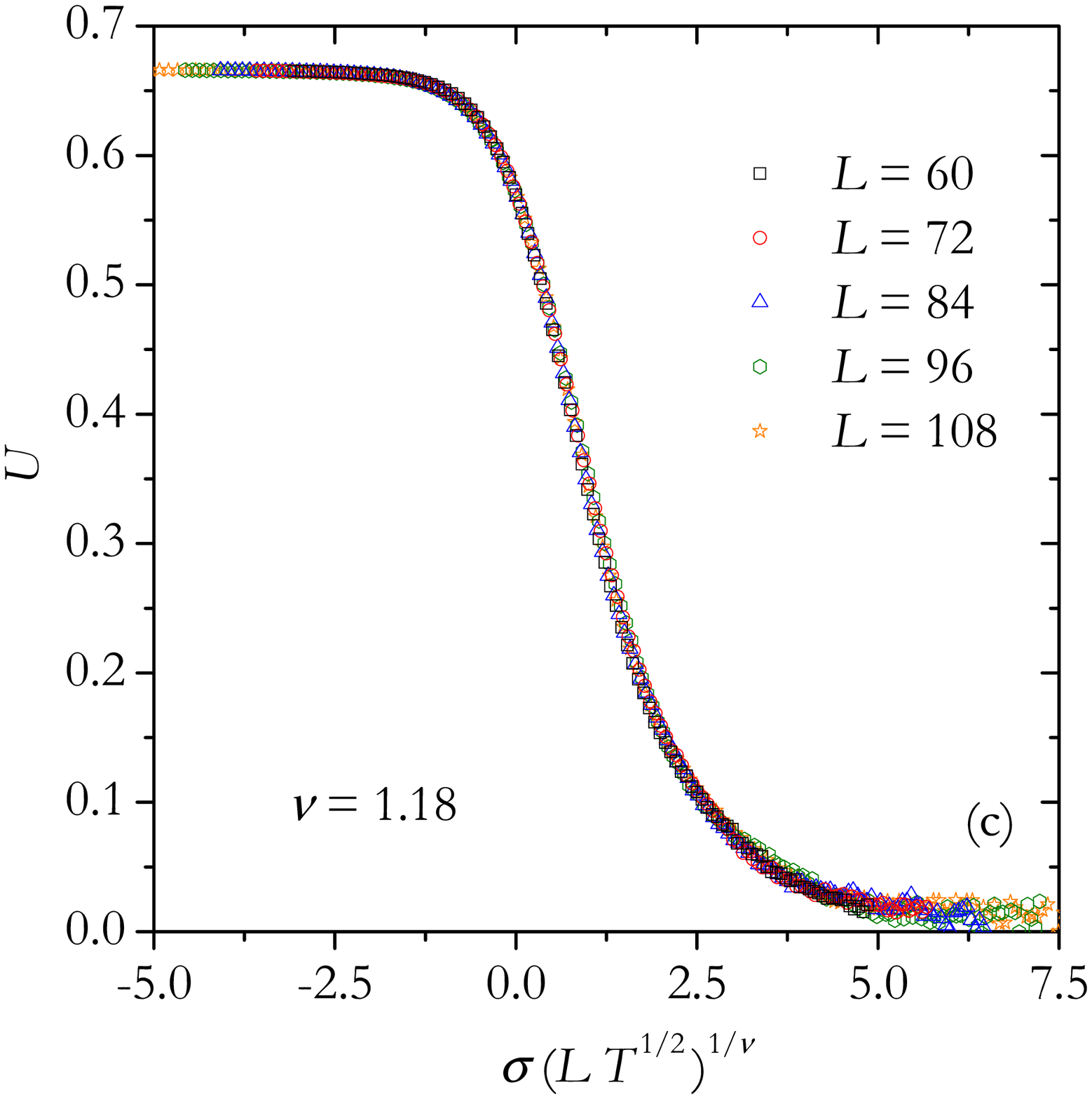}
\caption{(Color online) Same as Fig.~\ref{figure7} for the
extended scaling scheme.} \label{figure9}
\end{figure}

We start with the calculation of the order parameter
(Fig.~\ref{figure2}), susceptibility (Fig.~\ref{figure3}) and
cumulant (Fig.~\ref{figure4}) plotted versus $k_BT/w$ for several
lattice sizes. Due to computational limitations,~\cite{foot2} the
curves in Fig.~\ref{figure2} do not clearly show the existence, at
thermodynamic limit, of a finite temperature below which the order
parameter is different from zero. In order to clarify this point,
the inset in Fig.~\ref{figure2} shows the dependence of $\langle
\delta \rangle$ on $L^{-1}$ for constant $T$. As it can be
observed, $\langle \delta \rangle$ tends to a finite value (zero)
for $T<T_c$ ($T>T_c$).

From the intersections of the curves in Fig.~\ref{figure4} one
gets the estimation of the critical temperature. In this case,
$k_BT_c/w=0.182(1)$, which is in good agreement with the value
previously reported in the literature~\cite{SURFSCI3}. In
Ref.~\onlinecite{SURFSCI3}, the critical temperature was obtained
from the peaks of the curves of the specific heat versus
temperature (at fixed coverage) and coverage (at fixed
temperature). A more rigorous study was not possible due to the
lack of an adequate order parameter. In the inset, the data are
plotted over a wider range of temperatures, exhibiting the typical
behavior of the cumulants in presence of a continuous phase
transition. With respect to the value of the cumulant at the
transition temperature, $U^*$, this quantity was calculated by
plotting $U^*(L)$ vs $L^{-1/\nu}$ ~\cite{Ferren}, where the value
of $U^*(L)$ was obtained by fixing $K$($\equiv w/k_BT$) at our
estimate for $K_c$($ \equiv w/k_BT_c$) and looking at the cumulant
there (this is not shown here for brevity). In the thermodynamic
limit we obtained $U^*=0.57(1)$.  This value is consistent with
the cumulant crossings shown in Fig.~\ref{figure4}.

In order to discard the possibility that the phase transition is a
first-order one, the energy cumulants [Eq.~(\ref{cume})] have been
measured. As it is well-known, the finite-size analysis of $U_E$
is a simple and direct way to determine the order of a phase
transition~\cite{Binder2,Challa,Vilmayr}. Fig.~\ref{figure5}
illustrates the energy cumulants plotted versus $k_BT/w$ for
different lattice sizes ranging between $L=24$ and $L=60$. The
values of the parameters used in the MC runs were $m=21$,
$n_1=10^5$, $n_2=10^5$, $n_{MCS}=3 \times 10^5$, $T_1=0.25$ and
$T_m=0.15$. As it is observed, $U_E$ has the characteristic
behavior of a continuous phase transition: the minima in the
cumulants tend to $2/3$ as the lattice size is increased. This
indicates that the latent heat is zero in the thermodynamic limit,
which reinforces the arguments given in the paragraphs above.

Next, the critical exponents will be calculated. As stated in
Refs.~\onlinecite{Ferren,Janke,Bin}, the critical exponent $\nu$
can be obtained by considering the scaling behavior of certain
thermodynamic derivatives with respect to the inverse temperature
$K$, for example, the derivative of the cumulant and the
logarithmic derivatives of $\langle \delta \rangle$ and $\langle
\delta^2 \rangle$. In Fig.~\ref{figure6}(a) we plot the maximum
value of these derivatives as a function of system size on a
log-log scale~\cite{foot3}. The results for $1/\nu$ from these
fits are given in the figure. Combining these three estimates we
obtain $\nu=1.16(3)$ (see Table II). Once we know $\nu$, the
critical exponent $\gamma$ can be determined by scaling the
maximum value of the susceptibility~\cite{Ferren,Janke}. Our data
for $\chi|_{\rm max}$ are shown in Fig.~\ref{figure6}(b). The
value obtained for $\gamma$ is indicated in the figure and listed
in Table II.

On the other hand, the standard way to extract the exponent ratio
$\beta/\nu$ is to study the scaling behavior of $\langle \delta
\rangle$ at the point of inflection, i. e., at the point where $d
\langle \delta \rangle / d K $ is maximal. Since these points
should scale as usual, $\left(K^{\langle \delta \rangle}_{\rm
inf}- K_c \right)L^{1/\nu} \equiv \epsilon L^{1/\nu} ={\rm
const}$, we expect~\cite{Janke}
\begin{equation}
{\langle \delta \rangle |}_{\rm inf}=L^{-\beta/\nu} \tilde
\delta(\epsilon L^{1/\nu}) \propto L^{-\beta/\nu}, \label{betanu}
\end{equation}
where ${\langle \delta \rangle |}_{\rm inf}$ is the value of
$\langle \delta \rangle$ at the point of inflection. In addition,
since the derivative with respect to $K$ picks up a factor
$L^{1/\nu}$ from the argument of the scaling function $\tilde
\delta$,
\begin{equation}
\left. \frac{d \langle \delta \rangle}{d K} \right \vert _{\rm
max}=L^{\left(-\beta/\nu + 1/\nu \right)} \tilde \delta'(\epsilon
L^{1/\nu}) \propto L^{\left(1-\beta \right)/\nu}. \label{betabe}
\end{equation}

The scaling of ${\langle \delta \rangle |}_{\rm inf}$ is shown in
Fig.~\ref{figure6}(b). The linear fit through all data points
gives $\beta^{\left({\langle \delta \rangle |}_{\rm
inf}\right)}=0.24(6)$. In the case of $d \langle \delta \rangle/d
K |_{\rm max}$ [see Fig.~\ref{figure6}(b)], the value obtained
from the fit is $\beta^{\left(d \langle \delta \rangle/d K |_{\rm
max}\right)}=0.20(8)$. Combining the two estimates, we obtain the
final value $\beta=0.22(6)$, which is indicated in the figure and
listed in Table II.

The  finite-size scaling
theory~\cite{Binder0,Privman,Privman1,Ferren} allows for various
efficient routes to estimate $T_c$ from MC data. One of this
method, which was used in Fig.~\ref{figure4}, is from the
temperature dependence of $U(T)$ for different lattice sizes. An
independent procedure to determine $T_c$ will be used in the
following analysis. The method relies on the extrapolation of the
positions $K_c(L)$ of the maxima of various thermodynamic
quantities, which scale with system size
like~\cite{Binder0,Privman,Privman1,Ferren}
\begin{equation}
K_c(L)=K_c(\infty)+{\rm const}.L^{-1/\nu}. \label{nunu}
\end{equation}
Fig.~\ref{figure6}(c) shows a plot of $K_c(L)$ vs. $L^{-1/\nu}$
for the maxima of the slopes of $\langle \delta \rangle$, $\left(
d \langle \delta \rangle / d K \right)_{\rm max} $; $U$, $\left( d
U / d K \right)_{\rm max} $; $\ln \langle \delta \rangle$, $\left(
d \ln \langle \delta \rangle / d K \right)_{\rm max} $; $\ln
\langle \delta^2 \rangle$, $\left( d \ln \langle \delta^2 \rangle
/ d K \right)_{\rm max} $, as well as of the susceptibility,
$\chi_{\rm max}$. The lines are fits of the data to
Eq.~(\ref{nunu}) with $\nu=1.16$. From extrapolation one obtains
$K^{(A)}_c(\infty)$ [or $k_BT^{(A)}_c(\infty)/w$] for the
different observables $A$. In this case,
$k_BT^{(U)}_c(\infty)/w=0.1824(12)$; $k_BT^{(\ln \langle \delta
\rangle)}_c(\infty)/w=0.1825(14)$; $k_BT^{(\ln \langle \delta^2
\rangle)}_c(\infty)/w=0.1825(16)$; $k_BT^{(\langle \delta
\rangle)}_c(\infty)/w=0.1820(20)$; and
$k_BT^{(\chi)}_c(\infty)/w=0.1823(5)$. Combining these estimates
we find a final value $k_BT_c/w=0.1823(7)$, which coincides,
within numerical errors, with the value calculated from the
crossing of the cumulants.

Strong corrections to scaling were observed to be present at small
lattices and we excluded $L<60$ data from the calculations. On the
other hand, the $L \geq 60$ data are not good enough to include
corrections to scaling in the estimation of the critical exponents
and critical temperature. Thus, the quoted errors in our results
do not include systematic errors due to corrections to scaling
that possibly could affect our data.

The scaling behavior can be further tested by plotting $\langle
\delta \rangle L^{\beta/\nu}$ vs $|\epsilon|L^{1/\nu}$, $\chi
L^{-\gamma/\nu}$ vs $\epsilon L^{1/\nu}$ and $U$ vs $|\epsilon
|L^{1/\nu}$ and looking for data collapsing. Using our best
estimates $k_BT_c/w=0.182$, $\nu=1.16$, $\beta=0.22$ and
$\gamma=1.98$, we obtain very satisfactory scaling as it is shown
in Fig.~\ref{figure7}. This study leads to independent controls
and consistency checks of the values of all the critical
exponents. The collapses in Fig.~\ref{figure7} were calculated by
following a conventional FSS scheme.

In the next, the analysis of Figs.~\ref{figure6} and \ref{figure7}
is repeated, this time applying the extended FSS scheme mentioned
before. As it is shown in Fig.~\ref{figure8}, the values obtained
for the critical temperature $k_BT_c/w=0.1822(7)$ and the critical
exponents $\nu=0.18(3)$, $\beta=0.23(6)$ and $\gamma=2.04(5)$ (see
Table II) are in excellent agreement with those calculated using
the conventional FSS. In addition, we apply the extended FSS
scheme to collapse the data by plotting $U$, $\langle \delta
\rangle$ and $\chi$ in terms of the variables $\sigma \equiv
1-T_c/T$ and $L\sqrt{T}$. The results are shown in
Fig.~\ref{figure9}. The behavior of the critical quantities for
the extended FSS scheme is consistent with the behavior observed
by following the conventional FSS scheme, which reinforces the
robustness of the scaling analysis introduced here.

The critical exponents obtained by conventional and extended FSS,
along with the fixed point value of the cumulants, $U^*=0.57(1)$,
obtained in Fig.~\ref{figure4}, suggest that the phase transition
occurring for repulsive dimers on square lattices at $2/3$
monolayer coverage belongs to a new universality class. We say
``suggest" because the lattice sizes studied here do not allow us
to exclude a more complex critical behavior, for example, the
presence of a tricritical point~\cite{BORO5,WILDING1,WILDING2}. In
order to elucidate this point, future studies on the whole phase
diagram, including variables as coverage and an external field
(coupled to the order parameter) breaking the orientational
geometry of the phase, will be carried out.

\begin{table}[t]
\caption{Critical exponents obtained from the fits shown in
Figs.~\ref{figure6} and \ref{figure8}.}
\begin{ruledtabular}
\begin{tabular}{ccc}
exponent & conventional  & extended   \\
         &     FSS       &    FSS     \\
\hline
$\nu$   & $1.16(3)$ & $1.18(3)$    \\
$\beta$ & $0.22(6)$ & $0.23(6)$  \\
$\gamma$& $1.98(7)$ & $2.04(5)$   \\
$\alpha$& $\cdots$ & $\cdots$  \\
\end{tabular}
\end{ruledtabular}
\end{table}

Finally, we will briefly refer to the specific heat exponent
$\alpha$. The ``roughness" of the curves of $C$ prevents a direct
determination of $\alpha$. However, the usual hyperscaling
relations inequalities of Rushbrooke, $\alpha+2\beta+\gamma \geq
2$, and Josephson, $d \nu + \alpha \geq 2$ (being $d$ the
dimension of the space), predict a negative specific heat exponent
$\alpha \approx -0.3$. This finding is consistent with the
preliminary results obtained in the present study. Namely, even
though the fluctuations in the energy are of the order of the
statistical errors in the simulation results (reason for which the
data are not shown here), it is possible to observe that the
maxima of the curves of the specific heat remain practically
constant as $L$ is increased. In other words, the specific heat
seems not to diverge on approaching the transition. However, the
present data do not allow us to be conclusive on this point and
more work is needed to clearly elucidate the specific heats
behavior of the model.

In summary, we have addressed the critical properties of repulsive
dimers on two-dimensional square lattices at $2/3$ coverage. The
results were obtained by using exchange MC simulations and FSS
theory. The choice of an adequate order parameter [as defined in
Eq.~(\ref{fi2})] along with the exhaustive study of FSS presented
here allow us $1)$ to confirm previous results in the
literature,~\cite{Phares,SURFSCI3} namely, the existence of a
continuous phase transition at $2/3$ coverage; $2)$ to calculate
the critical temperature characterizing this transition; and $3)$
to obtain the complete set of static critical exponents for the
reported transition. Though it is not possible to exclude the
existence of a more complex critical behavior, the results suggest
that the phase transition belongs to a new universality class.

\acknowledgments This work was supported in part by CONICET,
Argentina, under Grant No. PIP 6294 and the Universidad Nacional
de San Luis, Argentina, under the Grants Nos. 328501 and 322000.



\begin{thebibliography}{99}
\bibitem{foot1}
As in the case of the $(2k \times 2)$ ordered phase occurring for
linear $k$-mers on square lattices at $1/2$ monolayer coverage,
~\cite{PRB5} the phase transition at $2/3$ monolayer coverage is
accomplished by a breaking of the translational and orientational
symmetries.
\bibitem{foot2}
As it is well-known, MC simulations of $k$-mers at equilibrium are
very time consuming. Consequently, the finite-size scaling study
was carried out for lattice sizes up to $L=108$, with an effort
reaching almost the limits of our computational capabilities.
\bibitem{foot3}

In this paper, the temperature derivatives were taken by averaging
the slopes of two adjacent data points as follows:  $\frac{1}{2}
\left[\frac{y_{i+1}-y_{i}}{x_{i+1}-x_{i}} +
\frac{y_{i}-y_{i-1}}{x_{i}-x_{i-1}} \right]$. Error bars were
estimated by propagation of errors from the last equation. On the
other hand, the coordinates of each maximum were calculated by
fittting a three-order polynomial to a set of between 15 and 20
data points around this critical value. Polynomials of order 2 and
4 were also used as fitting functions and the results did not
change significantly.
\end{thebibliography}
\end{document}